\documentclass[showpacs,aps,prd,reprint,superscriptaddress,nofootinbib,longbibliography]{revtex4-2}
\usepackage[colorlinks=true, pdfstartview=FitV,
bookmarks=true, bookmarksnumbered=true, breaklinks]{hyperref}
\usepackage[dvipdfmx]{graphicx}
\usepackage{amsmath,amssymb,bm,color,longtable,mathrsfs,slashed,comment,tikz}
\usepackage{ulem}
\usepackage{braket}
\usepackage{subfigure}

\definecolor{darkviolet}{rgb}{0.58, 0.0, 0.83}
\definecolor{electricultramarine}{rgb}{0.25, 0.0, 1.0}
\definecolor{brightpink}{rgb}{1.0, 0.0, 0.5}
\hypersetup{linkcolor=brightpink, citecolor=darkviolet, urlcolor=electricultramarine}

\definecolor{lime}{HTML}{A6CE39}
\DeclareRobustCommand{\orcidicon}{
	\hspace{-3mm}
	\begin{tikzpicture}
	\draw[lime, fill=lime] (0,0) 
	circle [radius=0.16] 
	node[white] {{\fontfamily{qag}\selectfont \tiny ID}};
	\draw[white, fill=white] (-0.0625,0.095) 
	circle [radius=0.007];
	\end{tikzpicture}
	\hspace{-3mm}
}

\foreach \x in {A, ..., Z}{\expandafter\xdef\csname orcid\x\endcsname{\noexpand\href{https://orcid.org/\csname orcidauthor\x\endcsname}
			{\noexpand\orcidicon}}
}

\begin{document}

\title{Scalar-glueball-mediated scale-anomaly dominance of the confining pressure of the pion in holographic QCD}

\author{Daisuke~Fujii\orcidB{}}
\email[]{daisuke@rcnp.osaka-u.ac.jp}
\affiliation{Advanced Science Research Center, Japan Atomic Energy Agency (JAEA), Tokai, 319-1195, Japan}
\affiliation{Research Center for Nuclear Physics, Osaka University, Ibaraki 567-0048, Japan}

\author{Akihiro~Iwanaka\orcidC{}}
\email[]{akihiro.iwanaka@yukawa.kyoto-u.ac.jp}
\affiliation{Yukawa Institute for Theoretical Physics, Kyoto University, Kyoto 606-8502, Japan}
\affiliation{Research Center for Nuclear Physics, Osaka University, Ibaraki 567-0048, Japan}

\author{Mitsuru Tanaka\orcidA{}}
\email[]{tanaka@hken.phys.nagoya-u.ac.jp}
\affiliation{Department of Physics, Nagoya University, Nagoya 464-8602, Japan}

\begin{abstract}
In this work, we analyze the energy density and the stress distribution inside the pion derived from gravitational form factors in top-down holographic QCD. In particular, we show that the dominance of the scale anomaly in the confining pressure, previously observed in the instant form for the nucleon, also holds for the pion in the light-front form. Furthermore, we find that in large-$N_c$ QCD described by this approach, the scalar glueball plays a mediating role in transmitting the confining pressure. These findings support the universal role of the scale anomaly in the stability of hadrons. 
\end{abstract}

\maketitle

\section{Introduction}

Quantum Chromodynamics (QCD) possesses classical scale invariance in the chiral limit, but this symmetry is broken at the quantum level due to renormalization effects, giving rise to a scale anomaly. This anomaly appears in the non-vanishing trace of the energy–momentum tensor (EMT). In the forward limit, the hadronic matrix element of the trace corresponds to the invariant mass for hadrons. It is well established that a significant fraction of the hadron mass originates from the anomalous contribution. Understanding the precise role of the scale anomaly in hadron formation is therefore a fundamental issue in hadron physics.

The matrix elements of the EMT in hadrons contain rich information beyond the mass. The corresponding form factors, i.e., gravitational form factors (GFFs) encode internal properties such as energy density, spin, pressure, and shear force distributions. By decomposing the EMT into its trace and traceless components, one can analyze how the scale anomaly contributes to each physical quantity. This framework also allows for the decomposition of hadron mass and spin~\cite{Ji:1994av,Ji:1996ek}. 
Notably, the 2018 experimental determination of the proton’s pressure distribution~\cite{Burkert:2018bqq} has opened a new avenue for exploring the internal mechanical properties of hadrons, stimulating active research from both theoretical~\cite{Polyakov:1999gs,Brommel:2005ee,Brommel:2007zz,LHPC:2007blg,Broniowski:2008hx,Frederico:2009fk,Masjuan:2012sk,Yang:2014xsa,Son:2014sna,Bali:2016wqg,Fanelli:2016aqc,Hudson:2017xug,Shanahan:2018pib,Alexandrou:2018xnp,Alexandrou:2019ali,Freese:2019bhb,Alexandrou:2020sml,Krutov:2020ewr,Shuryak:2020ktq,Pefkou:2021fni,Loffler:2021afv,deTeramond:2021lxc,Raya:2021zrz,Tong:2021ctu,Hackett:2023rif,Hackett:2023nkr,Liu:2023cse,HadStruc:2024rix,Xu:2023izo,Li:2023izn,Broniowski:2024oyk,Liu:2024jno,Liu:2024vkj,Hu:2024edc,Krutov:2024adh,Broniowski:2024mpw,Wang:2024lrm,Sugimoto:2025btn,Hatta:2025ryj,Goharipour:2025lep,Dehghan:2025ncw,Goharipour:2025yxm,Dehghan:2025eov,Cao:2025dkv,Ji:2025gsq,Fujii:2025pkv,Ji:2025qax} and experimental~\cite{Belle:2012wwz,Belle:2015oin,Kumano:2017lhr,Burkert:2018nvj,AbdulKhalek:2021gbh,Anderle:2021wcy,Duran:2022xag,Krutov:2022zgg,JeffersonLabHallA:2022pnx,CLAS:2022syx} perspectives.

In this paper, following Ref.~\cite{Fujii:2025aip}, we clarify the role of the scale anomaly in the pressure distribution of the pion using the trace–traceless decomposition. 
In our previous work~\cite{Fujii:2024rqd}, we computed the GFFs of the pion using a top-down holographic QCD model based on the D4–D8–$\overline{\text{D8}}$ configuration, so-called the Sakai–Sugimoto (SS) model~\cite{Sakai:2004cn, Sakai:2005yt}, which has successfully reproduced the hadron spectrum and form factors in Refs.~\cite{Constable:1999gb,Brower:2000rp,Sakai:2004cn,Hata:2007mb,Abidin:2008hn,Abidin:2008ku,Abidin:2009hr,Hashimoto:2009hj,Hashimoto:2009st,Imoto:2010ef,Ishii:2010ib,Liu:2017xzo,Liu:2017frj,Hashimoto:2019wmg,Liu:2019yye,Nakas:2020hyo,Hayashi:2020ipd,Fujii:2020jre,Suganuma:2020jng,Liu:2021tpq,Liu:2022urb,Sakai:2005yt,Hashimoto:2007ze,Hashimoto:2008zw,Hata:2008xc,Kim:2008pw,Grigoryan:2009pp,BallonBayona:2009ar,Bayona:2009pk,BallonBayona:2010ae,Cherman:2011ve,Bayona:2011xj,Harada:2011ur,Brunner:2015oqa,Li:2015oza,Druks:2018hif,Mamo:2019mka,Chakrabarti:2020kdc,Fujii:2021tsw,Liu:2021ixf,Liu:2021efc,Mamo:2021krl,Iwanaka:2022uje,Fujii:2022yqh,Mamo:2022eui,Fujii:2023ajs,Bigazzi:2023odl,Hechenberger:2023ljn,Ramalho:2023hqd,Allahverdiyeva:2023fhn,Castellani:2024dxr,Hechenberger:2024piy,Fujita:2022jus,Nasibova:2025wnw}. In the present study, we utilize the results in Ref.~\cite{Fujii:2024rqd} to evaluate the energy and stress distributions of the massless pion, focusing on how the scale anomaly governs the pressure profile via the trace–traceless decomposition. 
We also reconfirm that the EMT matrix element is dominated by three glueball states and newly provide a detailed breakdown of each state's individual contribution, which was not explored in Ref.~\cite{Fujii:2024rqd}.

In the instant-form formalism, spatial densities defined via three-dimensional Fourier transforms in the Breit frame inevitably depend on the choice of the constant instant time hypersurface. Lorentz boosts alter this surface, introducing ambiguities in defining the density for a given physical state~\cite{Miller:2018ybm,Jaffe:2020ebz}. In contrast, the light-front (LF) formalism, defined on constant LF time hypersurfaces, naturally fixes the center-of-mass motion and allows for an unambiguous definition of transverse densities via two-dimensional Fourier transforms~\cite{Burkardt:2000za,Burkardt:2002hr,Diehl:2002he,Miller:2018ybm,Jaffe:2020ebz,Freese:2021czn,Freese:2021mzg,Freese:2021qtb,Cao:2023ohj,Freese:2023abr}.
This is particularly relevant for systems like the pion, where the Compton wavelength is comparable to its spatial size and relativistic corrections become significant, as discussed in Ref.~\cite{Miller:2018ybm,Jaffe:2020ebz}. For such systems, LF transverse densities offer a unique and well-defined description of internal structure.

In this work, using the holographic QCD approach, we have shown that the scale anomaly generates a negative contribution to the pressure distribution inside the pion, dominated by scalar glueball. Combined with our earlier findings for the nucleon in instant-form ~\cite{Fujii:2025aip}, this implies that scale-anomaly–induced confining pressure is a universal mechanism, common across both forms (instant vs. light-front) and hadron types (nucleon vs. pion). These results offer valuable guidance for future theoretical and experimental studies of hadronic internal dynamics.

\section{Spatial distribution for pions in LF form}

The matrix elements of the EMT for pions are characterized by two independent GFFs, $A(t)$ and $D(t)$, and are defined as
\begin{align}
&\braket{\pi^a(p')|T^{\mu\nu}(0)|\pi^b(p)} \notag \\
&= \delta^{ab} \left[ 2P^\mu P^\nu A(t) + \frac{1}{2}(\Delta^\mu\Delta^\nu - \eta^{\mu\nu} \Delta^2) D(t) \right], \label{Def:GFFs}
\end{align}
where $P^\mu = (p^\mu + p'^\mu)/2$, $\Delta^\mu = p'^\mu-p^\mu$, $t = \Delta^2$, and $m_\pi$ is the pion mass. Throughout this paper, we use the mostly-plus metric $\eta^{\mu\nu} = \text{diag}(-1, +1, +1, +1)$, and adopt the normalization convention
$\langle \vec{p}^{\,\prime}
 | \vec{p} \rangle = 2p^0 (2\pi)^3 \delta^{(3)}(\vec{p}-\vec{p}^{\,\prime}
)$ with $p^0 = \sqrt{\vec{p}^2 + m_\pi^2}$.
Poincaré symmetry requires that $A(0) = 1$ in the forward limit $t=0$. In contrast, $D(t)$ is unconstrained at $t = 0$. It has been shown~\cite{Perevalova:2016dln} that the D-term is negative for stable systems. 

In our previous work~\cite{Fujii:2024rqd}, we computed the pion’s GFFs in the chiral limit using a top-down holographic QCD model based on the D4--D8--$\overline{\text{D8}}$ configuration. In the present study, we utilize these GFFs to analyze the spatial distributions of the pion.

It is common to define hadron density distributions as the three-dimensional Fourier transforms of form factors in the Breit frame, evaluated on equal instant time hypersurfaces ($x^0 = \mathrm{const.}$) in the instant form.
Because form factors are evaluated between plane-wave momentum eigenstates, the resulting spatial densities assume exact momentum localization. These states are completely delocalized in position space, so their Fourier transforms necessarily diverge~\cite{Miller:2018ybm}, casting doubt on the physical interpretation of such three-dimensional densities. To address this issue, a wave-packet approach has been proposed, in which the hadron’s center-of-mass is localized by smearing the state with a finite-width wave packet~\cite{Miller:2018ybm,Jaffe:2020ebz}. This construction reduces to the Fourier transform of form factors in the non-relativistic limit, while providing a relativistically consistent framework for defining spatial densities. For hadrons such as the pion—whose Compton wavelength is comparable to its spatial extent—these relativistic effects are especially important and cannot be neglected. Moreover, even when the center-of-mass is fixed using wave packets in the instant form, Lorentz boosts mix equal-time hypersurfaces, leaving residual frame dependence in the resulting densities.

To simultaneously address the above issues, we define two-dimensional transverse density distributions on constant LF time hypersurfaces ($x^+$: const.), i.e. on the transverse $x_\perp$ plane, where $x^\pm \equiv (x^0 \pm x^3)/\sqrt{2}$ and $x^i_\perp \equiv (x^1, x^2)$, following Refs.~\cite{Burkardt:2000za,Burkardt:2002hr,Diehl:2002he,Miller:2018ybm,Jaffe:2020ebz,Freese:2021czn,Freese:2021mzg,Freese:2021qtb,Cao:2023ohj,Freese:2023abr}. On this hypersurface, the longitudinal ($x^-$) direction remains delocalized as a plane wave, while the transverse $x_\perp$ plane can be localized using wave packets. This enables spatial distributions to be analyzed with a well-defined physical meaning. Moreover, in the LF formalism, a two-dimensional Galilean subgroup of the Poincaré group emerges, with the longitudinal momentum $P^+ \equiv \int dx^- d^2x_\perp T^{++} = (P^0 + P^3)/\sqrt{2}$ as its central charge. This structure guarantees the complete elimination of contamination from different time slices. Exploiting these advantages, we investigate the two-dimensional density distributions of pions on constant LF time hypersurfaces.

From Refs.~\cite{Burkardt:2002hr,Diehl:2002he,Miller:2018ybm,Freese:2021czn,Freese:2021mzg,Freese:2021qtb,Cao:2023ohj,Freese:2023abr}, the LF energy and stress distributions are defined as follows:
\begin{align}
&\epsilon_{\rm 2D}(x_\perp) \notag \\
&\equiv \int \frac{d^2 \Delta_\perp}{(2\pi)^2} e^{-i \vec{\Delta}_\perp \cdot \vec{x}_\perp} \frac{\braket{P^+, \vec{\Delta}_\perp/2 | T^{++} | P^+, -\vec{\Delta}_\perp/2}}{2P^+} \notag \\
&= P^+ \int \frac{d^2 \Delta_\perp}{(2\pi)^2} e^{-i \vec{\Delta}_\perp \cdot \vec{x}_\perp} A(t), \label{2Depsilon} \\
&S_{\rm 2D}^{ij}(x_\perp)  \notag \\
&\equiv \int \frac{d^2 \Delta_\perp}{(2\pi)^2} e^{-i \vec{\Delta}_\perp \cdot \vec{x}_\perp} \frac{\braket{P^+, \vec{\Delta}_\perp/2 | T^{ij} | P^+, -\vec{\Delta}_\perp/2}}{2P^+} \notag \\
&= -\frac{1}{4P^+} \int \frac{d^2 \Delta_\perp}{(2\pi)^2} e^{-i \vec{\Delta}_\perp \cdot \vec{x}_\perp} (\Delta^i_\perp \Delta^j_\perp - \Delta_\perp^2 \delta^{ij}) D(t), \label{2Dstress}
\end{align}
with $x_\perp=|\vec{x}_\perp|$, $\vec{\Delta}_\perp=(\Delta^1,\Delta^2)$ and $i,j=1,2$. The stress tensor can be expressed using the transverse pressure $p_{\rm 2D}$ and shear force $s_{\rm 2D}$ as
\begin{align}
S_{\rm 2D}^{ij} = \delta^{ij} p_{\rm 2D} + \left( \frac{x^i_\perp x^j_\perp}{x_\perp^2} - \frac{1}{2} \delta^{ij} \right) s_{\rm 2D}. \label{Spherical2Dstress}
\end{align}

Decomposing the EMT into its trace $\hat{T}^{\mu\nu} \equiv (g^{\mu\nu}/4) {T^\rho}_\rho$ and traceless components $\bar{T}^{\mu\nu} \equiv T^{\mu\nu} - (g^{\mu\nu}/4){T^\rho}_\rho$, 
the pressure can be separated into two parts:
\begin{align}
&p_{\rm 2D} = \bar{p}_{\rm 2D} + \hat{p}_{\rm 2D}, \label{decomp_pressure} \\
&\bar{p}_{\rm 2D}(x_\perp)=\frac{1}{2P^+} \int \frac{d^2 \Delta_\perp}{(2\pi)^2} e^{-i \vec{\Delta}_\perp \cdot \vec{x}_\perp} \notag \\
&\hspace{26mm}\times\left[ \frac{m_\pi^2}{2} A(t) + \frac{\Delta_\perp^2}{8} (A(t) + D(t)) \right], \notag \\
&\hat{p}_{\rm 2D}(x_\perp)=\frac{1}{2P^+} \int \frac{d^2 \Delta_\perp}{(2\pi)^2} e^{-i \vec{\Delta}_\perp \cdot \vec{x}_\perp} \notag \\
&\hspace{26mm}\times\left[ -\frac{m_\pi^2}{2} A(t) - \frac{\Delta_\perp^2}{8} (A(t) + 3D(t)) \right], \notag
\end{align}
where $\hat{p}_{\rm 2D}$ corresponds to the contribution from the trace anomaly.

From these distributions, the mass radius $\langle x_\perp^2 \rangle_{\text{mass}}$ and mechanical radius $\langle x_\perp^2 \rangle_{\text{mech}}$ are defined as
\begin{align}
&\langle x_\perp^2 \rangle_{\text{mass}} = \frac{\int d^2x_\perp  x_\perp^2 \epsilon_{2D}(x_\perp)}{\int d^2x_\perp  \epsilon_{2D}(x_\perp)} = 4 \left. \frac{dA(t)}{dt} \right|_{t=0}, \\
&\langle x_\perp^2 \rangle_{\text{mech}} = \frac{\int d^2x_\perp x_\perp^2 (x^i_\perp x^j_\perp/x_\perp^2) S^{ij}_{\rm 2D}(x_\perp)}{\int d^2x_\perp (x^i_\perp x^j_\perp/x_\perp^2) S^{ij}_{\rm 2D}(x_\perp)} \notag \\
&\hspace{14mm}= \frac{4D(0)}{\int_{-\infty}^{0} dt D(t)},
\end{align}
where $(x^i_\perp x^j_\perp / x_\perp^2) S^{ij}_{\rm 2D}$ represents the pressure acting in the radial direction.

\section{Gravitational from factors from top-down holographic QCD}

In this section, we formulate a method for evaluating the GFFs of the pion within the top-down holographic QCD approach~\cite{Sakai:2004cn,Sakai:2005yt}, by applying the framework previously developed for the nucleon in Ref.~\cite{Fujita:2022jus}.

The SS model, employed in this work as a top-down holographic QCD framework, is constructed by embedding $N_f$ D8-branes as probes into the background geometry generated by $N_c$ D4-branes~\cite{Witten:1998zw}. This geometry corresponds to a ten-dimensional curved spacetime in type IIA supergravity, which is derived from eleven-dimensional supergravity (M theory) through compactification on a doubly Wick-rotated anti-de Sitter space in seven dimensions ($\rm AdS_7$) black hole background times a four-sphere:
\begin{align}
    ds^2=& \frac{r^2}{L^2}\big[f(r)d\tau^2-dx^2_0+dx_1^2+dx^2_2+dx^2_3+dx^2_{11}\big] \notag \\ 
    &+\frac{L^2}{r^2}\frac{dr^2}{f(r)}+\frac{L^2}{4}d\Omega^2_4 , \label{AdS7BH} \\
    f(r)=& 1-\frac{R^6}{r^6}, \ \ \ R = \frac{L^2M_{\rm KK}}{3}, \ \ \ L^3=8\pi g_s N_c l^3_s, \ \ \ \notag \\
    R_{11}=&
    \frac{\lambda}{2\pi N_c M_{\rm KK}}, \ \ \ \lambda N_c=\frac{L^6M_{\rm KK}}{32\pi g_s l^5_s}=216\pi^3\kappa, \notag
\end{align}
where $M_{\rm KK}$ denotes the Kaluza–Klein mass scale associated with the $\tau$ direction, $g_s$ is the string coupling, $l_s$ is the string length, $\lambda$ is the 't Hooft coupling, and $R_{11}$ represents the radius of the compactified $x_{11}$ dimension.

The boundary expectation values of the EMT, $\big<T_{\mu\nu}\big>$, can be obtained by analyzing the asymptotic behavior of the bulk metric perturbations in the background geometry given by Eq.~\eqref{AdS7BH}. Specifically, the relation is expressed as
\begin{align}
    \int_{x_{11},\tau}\delta g_{\mu\nu}\sim\frac{2\kappa^2_7 L^5}{6}\frac{\big<T_{\mu\nu}\big>}{r^4}+..., \label{boundaryEMT}
\end{align}
where $\kappa^2_7$ is connected to the 11-dimensional gravitational constant through $\kappa^2_{11}={\rm Vol}(S^4)\kappa^2_7=8\pi^2/3(L/2)^4\kappa^2_7$. 
To derive this result, one must employ holographic renormalization within the framework of M-theory, as formulated in Ref.~\cite{deHaro:2000vlm}. The advantage of working in eleven dimensions lies in the technical simplification of the renormalization procedure.\footnote{
Similar methods for non-conformal 10-dimensional backgrounds have also been established~\cite{Kanitscheider:2008kd}.
}

To obtain the metric fluctuations $\delta g_{MN}$ induced by the pion field, we solve the linearized Einstein equations in the bulk spacetime. These equations take the form:
\begin{align}
    \bar{\mathcal{H}}_{MN}=&\mathcal{H}_{MN}-\frac{g_{MN}}{2}\mathcal{H}^P_P=-2\kappa^2_7\mathcal{T}_{MN}, \label{linearEq} \\
    \mathcal{H}_{MN}\equiv& 
    \nabla^2\delta g_{MN}+\nabla_M\nabla_N\delta g^ P_P \notag \\
    &-\nabla^P(\nabla_M\delta g_{NP}+\nabla_N\delta g_{MP})-\frac{12}{L^2}\delta g_{MN}, \notag
\end{align}
where $M,N,P = 0,1,2,3,\tau,r,11$ label the coordinates of the seven-dimensional spacetime obtained after compactifying $S^4$. From this equation, we derive the conservation law of the bulk EMT, $\nabla^M\mathcal{T}_{MN}=0$. The source term $\mathcal{T}_{MN}$ represents the EMT of the bulk matter, which, in our analysis, arises solely from the pion field. 

To evaluate the bulk EMT $\mathcal{T}_{MN}$, we consider the embedding of D8 branes into the eleven-dimensional background, and write the D8-brane action as
\begin{align}
    &S_{D8}= \notag \\
    &-\frac{C}{2\pi R_{11}g_{s}}\int d\tau dx_{11}d^4xdrd\Omega_4(\delta(\tau)+\delta(\tau-\pi/M_{\rm KK})) \notag \\
    &\hspace{10mm}\times\frac{\sqrt{-G_{(11)}}}{\sqrt{G_{\tau\tau}}}\frac{1}{4}G^{MN}G^{PQ}{\rm tr}[F_{MP}F_{NQ}]+...,
\end{align}
with $C=(64\pi^6l_s^5)^{-1}$ and the background metric $G_{MN}$ from Eq.~\eqref{AdS7BH}. The EMT is defined as
\begin{align}
    \mathcal{T}_{MN}\equiv-\frac{2}{\sqrt{-G_{(11)}}}\frac{\delta S_{D8}}{\delta G^{MN}_{(11)}}{\rm Vol}(S^4) , \notag
\end{align}
which gives the following expressions:
\begin{align}
    &\mathcal{T}_{MN}(\vec{\Delta},r) \equiv 2\pi R_{11}\int d\tau^\prime \mathcal{T}_{MN}(\tau^\prime,\vec{\Delta},r), \label{bulkEMT} \\
    &\mathcal{T}_{\mu\nu} (\vec{\Delta},r)= \notag \\
    &\frac{L^5}{{r}^5}\frac{\pi^2C{r}^2 L^2}{3g_s \sqrt{f(r))}}{\rm tr}\Big[F_{\mu\rho}{F_\nu}^{\rho}+k(z)^{4/3}M_{\rm KK}^2F_{\mu z}F_{\nu z} \notag \\
    &-\frac{\eta_{\mu\nu}}{4}\big(F_{\rho\sigma}F^{\rho\sigma}+2k(z)^{4/3}M^2_{\rm KK}F_{\rho z}{F^\rho}_z\big)\Big], \notag \\
    &\mathcal{T}_{11,11}(\vec{\Delta},r)= \notag \\
    &\frac{L^5}{{r}^5}\frac{\pi^2 Cr L^2}{3g_s\sqrt{f(r)}}{\rm tr}\Big[-\frac{1}{4}F_{\rho\sigma}F^{\rho\sigma}-\frac{k(z)^{4/3}M^2_{\rm KK}}{2}F_{\rho z}F^{\rho z}\Big], \notag \\
    &\mathcal{T}_{\mu z}(\vec{\Delta},r)=\frac{L^5}{{r}^5}\frac{\pi^2 Cr L^2}{3g_s\sqrt{f(r)}}{\rm tr}\Big[F_{\mu\rho}{F_z}^\rho\Big], \notag \\ 
    &\mathcal{T}_{zz}(\vec{\Delta},r)= \notag \\
    &\frac{L^5}{{r}^5}\frac{\pi^2 Cr L^2}{3g_s\sqrt{f(r)}}{\rm tr}\Big[\frac{1}{2}F_{z\rho}{F_z}^\rho-\frac{1}{4k(z)^{4/3}M^2_{\rm KK}}F_{\rho\sigma}F^{\rho\sigma}\Big], \notag
\end{align}
where $r$ and $z$ are related via $z=\pm\sqrt{r^6/R^6-1}$ for $r\in[R,\infty)$, and $k(z)=1+z^2$. 
In our setup, the pion is treated as the only bulk field generating gravitational perturbations. Then, the gauge field $A_M(x^\nu,z)$ is decomposed as
\begin{align}
    &A_\mu(x^\nu,z)=0, \ \ \  A_z(x^\nu,z)=\pi(x^\mu)\phi_0(z), \label{pion}\\
    &\phi_0(z)=c_0k(z)^{-1}, \notag
\end{align}
where $c_0^{-2}=2M^2_{\rm KK}\kappa\int dzk(z)\phi_0(z)^2$ ensures normalization~\cite{Sakai:2004cn}. Since the pion field appears as external lines, it is represented by a plane wave with incoming momentum $p_\nu$, as $\pi(x^\nu) = e^{ip_\nu x^\nu}$. 

The metric fluctuation $\delta g_{\mu\nu}$ relevant to our analysis is required to be transverse $\Delta^\mu\delta g_{\mu\nu}\sim0$ near the boundary, because the conservation of the EMT is satisfied. Such a solution can be written as $\delta g_{ij}\sim\delta g^{\rm TT}_{ij}+\frac{1}{5}(\delta_{ij}-\Delta_i\Delta_j/\vec{\Delta}^2)\delta g^\alpha_\alpha$, where $\delta g^{\rm TT}_{ij}$ is the transverse-traceless (TT) part with $i,j=1,2,3$ and $\alpha=\mu,11,\tau$. Furthermore, in the asymptotically AdS boundary ($r\rightarrow\infty$) labeled by $(x^\mu,\tau,x^{11})$ of M-theory, the six-dimensional traceless condition $\delta g^\alpha_\alpha\sim0$ is required. Therefore, we only need to solve the TT modes of the Einstein equation under the axial gauge, which are nothing but the QCD glueball spectra. These TT modes consist of 14 independent components, each of which can be assigned a definite spin quantum number~\cite{Brower:2000rp} (see also Ref.~\cite{Constable:1999gb}). Since the background spacetime possesses an SO(4) symmetry in the $x^1\text{–}x^3$ and $x^{11}$ directions, the TT modes can be decomposed into irreducible representations of SO(4): a 9-, 4-, and a 1-dimensional block. Each of these blocks satisfies the same linearized Einstein equations. Furthermore, by decomposing these under the SO(3) symmetry of the $x^1\text{–}x^3$ spatial directions, the components can be classified according to their spin. Among the 14 TT modes, only three contribute to the matrix element of EMT: the spin-2 and spin-0 components in the 9-dimensional representation of SO(4), denoted as ${\rm T}_4(2^{++})$ and ${\rm T}_4(0^{++})$, and the spin-0 component in the 1-dimensional representation, denoted as ${\rm S}_4(0^{++})$.
\footnote{${\rm T}_4(0^{++})$ state is the counterpart of the dilaton in the bulk theory, while ${\rm S}_4(0^{++})$ state is the lightest glueball state in this model. The latter has a nonzero component along the compact $\tau$ direction and is therefore referred to as the exotic scalar state in Ref.~\cite{Brower:2000rp}. The corresponding boundary gluonic operators are given in Ref.~\cite{Brunner:2015oqa} as
\begin{align}
&\mathcal{O}_D=+\frac{3}{8}\mathrm{Tr}G^{\mu\nu}G_{\mu\nu}+\cdots, \ \ \ \rm [for \ {\rm T}_4(0^{++})]\notag \\
&\mathcal{O}_E=-\frac{5}{8}\mathrm{Tr}G^{\mu\nu}G_{\mu\nu}+\cdots,  \ \ \ \rm [for \ {\rm S}_4(0^{++})]\notag
\end{align}
so that $\mathcal{O}_D-\mathcal{O}_E=\mathrm{Tr}G^{\mu\nu}G_{\mu\nu}$ identifies the scalar glueball operator in QCD, 
where $G_{\mu\nu}$ denotes the field strength of the color gauge field at the boundary. 
Since this linear combination is not an eigenmode on the gravity side, we keep both ${\rm T}_4(0^{++})$ and ${\rm S}_4(0^{++})$ modes in our analysis. We note that Ref.~\cite{Brunner:2015oqa} concludes that the ${\rm S}_4(0^{++})$ state is likely a holographic artifact, since its mass is smaller and its decay width larger than those of the lightest scalar glueball in QCD. 
 \label{footnote2}}
The remaining modes decay more rapidly near the boundary and do not contribute to the matrix element of EMT.\footnote{Actually, substituting the pion field \eqref{pion} into Eq.~\eqref{gOther}, we confirm that $\delta g^{\rm other}_{\mu\nu}$ falls faster than $\rm T_4$ and $\rm S_4$ mode at $r\rightarrow\infty$. }

Among the modes that contribute to the matrix element, we select the components that exhibit the same tensor structure as in Eq.~\eqref{Def:GFFs}:
\begin{align}
    &\delta g^{{\rm T}(2)}_{\alpha\beta}\big|_{r\rightarrow\infty}=\frac{2\kappa^2_7L^5}{6r^4}
    \begin{pmatrix}
    2\vec{\Delta}^2 & 0 & 0 & 0 \\
    0 & \vec{\Delta}^2\delta_{ij}-\Delta_i\Delta_j & 0 & 0 \\
    0 & 0 & 0 & 0 \\
    0 & 0 & 0 & 0
    \end{pmatrix}t_2(t) \label{T2++} \\
    &\delta g^{{\rm T}(0)}_{\alpha\beta}\big|_{r\rightarrow\infty}=\frac{2\kappa^2_7L^5}{6r^4}
    \begin{pmatrix}
    \Delta^2\eta_{\mu\nu}-\Delta_\mu\Delta_\nu & 0 & 0 \\
    0 & -3\Delta^2 & 0 \\
    0 & 0 & 0 
    \end{pmatrix}t_0(t) \label{T0++} \\
    &\delta g^{{\rm S}(0)}_{\alpha\beta}\big|_{r\rightarrow\infty}=\frac{2\kappa^2_7L^5}{6r^4}
    \begin{pmatrix}
    \Delta^2\eta_{\mu\nu}-\Delta_\mu\Delta_\nu & 0 & 0 \\
    0 & \Delta^2 & 0 \\
    0 & 0 & -4\Delta^2 
    \end{pmatrix}s_0(t) \label{S0++} ,
\end{align}
where the ${\rm T}_4(2^{++})$ mode is traceless in four-dimensional spacetime, while the ${\rm T}_4(0^{++})$ mode includes SO(4)-symmetric space components and corresponds to the dilaton in type IIA supergravity. The ${\rm S}_4(0^{++})$ mode is the only one that contains a nonzero $\tau$ component.

The expectation values of the temporal and spatial components of the EMT can then be written as
\begin{align}
    &\braket{T_{00}}=\frac{6r^4}{2\kappa^2_7L^5}\left(\delta g^{{\rm T}(2)}_{00}+\delta g^{{\rm T}(0)}_{00}+\delta g^{{\rm S}(0)}_{00}\right)\big|_{r\rightarrow\infty} \notag \\
    &=(2t_2-t_0-s_0)\vec{\Delta}^2 \label{temporal} \\
    &\braket{T_{ij}}=\frac{6r^4}{2\kappa^2_7L^5}\left(\delta g^{{\rm T}(2)}_{ij}+\delta g^{{\rm T}(0)}_{ij}+\delta g^{{\rm S}(0)}_{ij}\right)\big|_{r\rightarrow\infty} \notag \\
    &=(t_2+t_0+s_0)(\vec{\Delta}^2\delta_{ij}-\Delta_i\Delta_j).\label{spatial}
\end{align}

By decomposing the metric fluctuations as $\delta g_{\mu\nu}=\delta g^{\rm T(2)}_{\mu\nu}+\delta g^{\rm T(0)}_{\mu\nu}+\delta g^{\rm S(0)}_{\mu\nu}+\delta g^{\rm other}_{\mu\nu}$, the linearized Einstein equations are reduced to the following form:
\begin{align}
    &\frac{r^2}{L^2}\Big[\frac{1}{L^2r^5}\partial_r\big((r^7-rR^6)\partial_r\big)-\frac{L^2\Delta^2}{r^2}\Big]\frac{L^2}{r^2}\delta g^{\rm T}_{\mu\nu}=-2\kappa^2_7\mathcal{T}^{\rm T}_{\mu\nu}, \notag \\
    &\frac{r^2}{L^2}\Big[\frac{1}{L^2r^5}\partial_r\Big((r^7-rR^6)\Big(\partial_r+\frac{144r^5R^6}{(5r^5-2R^6)(r^6+2R^6)}\Big)\Big) \notag \\
    &\hspace{26mm}-\frac{L^2\Delta^2}{r^2}\Big]\frac{L^2}{r^2}\delta g^{\rm S(0)}_{\mu\nu}\Big|_{r\rightarrow\infty}=-2\kappa^2_7\mathcal{T}^{\rm S(0)}_{\mu\nu},\notag 
\end{align}
where the EMT is decomposed as $\mathcal{T}_{\mu\nu} = \mathcal{T}^{\rm T}_{\mu\nu} + \mathcal{T}^{\rm S(0)}_{\mu\nu} + \mathcal{T}^{\rm other}_{\mu\nu}$ with $\mathcal{T}^{\rm T}_{\mu\nu} \equiv \mathcal{T}^{\rm T(2)}_{\mu\nu} + \mathcal{T}^{\rm T(0)}_{\mu\nu}$. Since $\delta g^{\rm T(2)}_{\mu\nu}$ and $\delta g^{\rm T(0)}_{\mu\nu}$ satisfy the same equation, they are treated collectively as $\delta g^{\rm T}_{\mu\nu}$.

In line with the method described in Ref.~\cite{Fujita:2022jus}, each contribution to the EMT can be expressed as
\begin{align}
    &2\kappa^2_7\mathcal{T}^{\rm T}_{\mu\nu}=2\kappa^2_7\mathcal{T}_{\mu\nu}-2\kappa^2_7\mathcal{T}^{\rm S}_{\mu\nu}-2\kappa^2_7\mathcal{T}^{\rm other}_{\mu\nu}, \\
    &2\kappa^2_7\mathcal{T}^{\rm S}_{\mu\nu}= \notag \\
    &\frac{r^6+2R^6}{4(r^6-R^6)}\Big(\eta_{\mu\nu}-\frac{\Delta_\mu \Delta_\nu}{\Delta^2}\Big)\Big[\frac{1}{L^4r^3}\partial_r\big(r(r^6-R^6)\partial_r(a+b) \notag \\
    &\hspace{40mm}-3R^6(a+b)\big)-\Delta^2b\Big], \\ 
    &2\kappa^2_7\mathcal{T}^{\rm other}_{\mu\nu}= \notag \\
    &\frac{1}{L^4r^3}\partial_r\Big[\Big(\eta_{\mu\nu}-\frac{\Delta_\mu \Delta_\nu}{\Delta^2}\Big)r(r^6-R^6)\partial_r a-3R^6\eta_{\mu\nu}b\Big],  \\
    &\partial_r a=-\frac{3R^6}{(5r^6-2R^6)}b+\frac{9r^{11}}{(5r^6-2R^6)R^6}2\kappa^2_7\mathcal{T}_{zz}, \notag \\ 
    &b=-i\frac{\Delta^\mu}{\Delta^2}\frac{r^6\sqrt{r^6-R^6}}{R^9} 2\kappa^2_7 \mathcal{T}_{\mu z}, 
\end{align}
with the following ansatz:
\begin{align}
    &\delta g^{\rm other}_{\mu\nu}(\vec{\Delta},r)=\frac{r^2}{L^2}\frac{\Delta_\mu \Delta_\nu}{\Delta^2}a(r,\Delta), \notag \\
    &\delta g^{\rm other}_{11,11}(\vec{\Delta},r)=\frac{r^2}{L^2}b(r,\Delta), \notag \\
    &\delta g^{\rm other}_{\tau\tau}(\vec{\Delta},r)=\frac{r^2f(r)}{L^2}(-b(r,\Delta)). \label{gOther}
\end{align}

We obtain the metric perturbation sourced by the pion field in the bulk via the Green's function method as
\begin{align}
    &\delta g^{\rm T/S}_{\mu\nu}(\vec{\Delta},r)\sim\frac{r^2}{L^2}\int dr^\prime \sqrt{-G_{(7)}(r^\prime)}G^{\rm T/S}(\vec{\Delta},r,r^\prime) \notag \\
    &\hspace{40mm}\times\frac{L^2}{{r^\prime}^2}2\kappa^2_7\mathcal{T}^{\rm T/S}_{\mu\nu}(\vec{\Delta},r^\prime), \label{solg}
\end{align}
where for the $\rm S_4$ mode, this form holds asymptotically as $r\rightarrow\infty$, which is indicated by the symbol $\sim$. 
The Green’s function $G^{\rm T/S}$ is expanded in terms of the complete set ${\Psi^{\rm T/S}_n(r)}$ as
\begin{align}
    G^{\rm T/S}(\vec{\Delta},r,r^\prime)=\sum^\infty_{n=1}\frac{\Psi^{\rm T/S}_n(r)\Psi^{\rm T/S}_n(r^\prime)}{(m^{\rm T/S}_n)^2+\vec{\Delta}^2},
\end{align}
where $\Psi^{\rm T/S}_n(r)$ are solutions to the following eigenvalue problems:
\begin{align}
    &\frac{1}{r^3L^4}\partial_r\big((r^7-rR^6)\partial_r\Psi^{\rm T}_n\big)=-(m^{\rm T}_n)^2\Psi^{\rm T}_n, \\
    &\frac{1}{r^3L^4}\partial_r\Big((r^7-rR^6)\Big(\partial_r+\frac{144r^5R^6}{(5r^6-2R^6)(r^6+2R^6)}\Big)\Psi^{\rm S}_n\Big) \notag \\
    &\hspace{45mm}=-(m^{\rm S}_n)^2\Psi^{\rm S}_n, \label{EVeq}
\end{align}
where $m^{\rm T/S}_n$ denotes the $n$-th glueball mass for the respective modes. These eigenfunctions satisfy the orthonormality and completeness relations:
\begin{align}
    &\frac{r^3}{L^3}\sum^\infty_{n=1}\Psi^{\rm T/S}_n(r)\Psi^{\rm T/S}_n(r^\prime)=\delta(r-r^\prime), \notag \\
    &\int^\infty_R dr\frac{r^3}{L^3}\Psi^{\rm T/S}_n(r)\Psi^{\rm T/S}_m(r)=\delta_{mn}. \label{CR}
\end{align}
It is worth noting that, from Eq.~\eqref{T2++}-\eqref{S0++}, the dominant sources for the $\rm T_4(2^{++})$, $\rm T_4(0^{++})$, and $\rm S_4(0^{++})$ modes in the bulk are $\mathcal{T}_{\mu\nu}$, $\mathcal{T}_{11,11}$, and $\mathcal{T}_{\tau\tau}$, respectively.\footnote{Although the full $\mathcal{T}_{\tau\tau}$ vanishes due to Eq.~\eqref{bulkEMT}, the individual $\rm S_4$ contribution satisfies $\mathcal{T}_{\tau\tau}^{\rm S} + \mathcal{T}_{\tau\tau}^{\rm other} = 0$, indicating that $\mathcal{T}_{\tau\tau}^{\rm S}$ alone acts as the source relevant for the propagation of the $\rm S_4$ mode.}

From Eq.~\eqref{boundaryEMT}, the matrix elements of the EMT can be extracted by reading off the coefficients of the $1/r^4$ terms in the metric fluctuations $\delta g^{\rm T/S}_{\mu\nu}$. The result can be expressed as
\begin{align}
    &\braket{\pi^a(p^\prime)|T_{\mu\nu}|\pi^a(p)}=\braket{T^{\rm T}_{\mu\nu}}+\braket{T^{\rm S}_{\mu\nu}} \\
    &\braket{\pi^a(p^\prime)|T^{\rm T/S}_{\mu\nu}|\pi^a(p)} \notag \\
    &=\frac{6}{L^{10}}\sum_{n=1}^\infty\alpha_n^{\rm T/S}\int^\infty_Rdr^\prime\frac{{r^\prime}^3\Psi^{\rm T/S}(r^\prime)}{\vec{\Delta}^2+(m_n^{\rm T/S})^2}\mathcal{T}^{\rm T/S}_{\mu\nu}(\pi,\phi_0;r^\prime,\vec{\Delta}), \notag
\end{align}
with $\alpha^{\rm T/S}=(L^4/6)(m^{\rm T/S}_n)^2\int^\infty_R dr r^3\Psi^{\rm T/S}_n(r)$.\footnote{
In the axial gauge, the consevation law of the bulk EMT can be written as
\begin{align}
    &\partial^\mu\mathcal{T}_{\mu a}=-\frac{1}{L^4}{r^3}\partial_r(r(r^6-R^6)\mathcal{T}_{ra}), \notag \\
    &\eta^{ab}\mathcal{T}_{ab}=\frac{1}{L^4r^2}(L^4r^3\partial^\mu\mathcal{T}_{\mu r}+r(r^6-R^6)\partial_rT_{rr} \notag \\
    &\hspace{22mm}+(8r^6+R^6)\mathcal{T}_{rr})-\frac{r^6(r^6+2R^6)}{(r^6-R^6)^2}\mathcal{T}_{\tau\tau}, \notag
\end{align}
with $a,b=0,1,2,3,11$. Using the first line, we derive
\begin{align}
    &ik^\mu\mathcal{T}^{\rm T}_{\mu\nu}=ik^\mu(\mathcal{T}_{\mu\nu}-\mathcal{T}^{\rm S}_{\mu\nu}-\mathcal{T}^{\rm other}_{\mu\nu})=0 \notag
\end{align}
where $\partial^\mu\mathcal{T}^{\rm S}_{\mu\nu}$ vanishes by the tensor structure. From these relations, it is then clear that the matrix elements of the four-dimensional EMT are conserved.
} 
Here, the expression $\mathcal{T}^{\rm T/S}_{\mu\nu}(\pi, \phi_0; r', \vec{\Delta})$ denotes the bulk EMT evaluated using the pion field Eq.~\eqref{pion}. 
After changing to the $z$-coordinate, the matrix elements become
\begin{align}
    &\braket{\pi^a(p^\prime)|T_{\mu\nu}|\pi^a(p)}= \notag \\
    &\frac{1}{L^7}\int_{-\infty}^\infty dz\sum^\infty_{n=1}\frac{\alpha^{\rm T}_n\Psi^{\rm T}_n(r^\prime(z))}{(m^{\rm T}_n)^2+\vec{\Delta}^2}\Big[2P_\mu P_\nu\Big(J(z)\Big) \notag \\
    &+\frac{1}{2}(\Delta_\mu\Delta_\nu-\eta_{\mu\nu}\Delta^2)\Big(-J(z)-J^{\rm S}(z)-J^{\rm other}(z)\Big)\Big] \notag \\
    &+\frac{1}{L^7}\int_{-\infty}^\infty dz\sum^\infty_{n=1}\frac{\alpha^{\rm S}_n\Psi^{\rm S}_n(r^\prime(z))}{(m^{\rm S}_n)^2+\vec{\Delta}^2}\Big[\frac{1}{2}(\Delta_\mu\Delta_\nu-\eta_{\mu\nu}\Delta^2)J^{\rm S}(z)\Big]
\end{align}
where the source $J(z)$, $J^{\rm S}(z)$, and $J^{\rm other}(z)$ are given by
\begin{align}
    &J(z)=12\kappa(1+z^2)\phi_0(z)^2 \notag \\
    &J^{\rm S}(z)=9\kappa(1+z^2)(3+z^2)\phi_0(z) \notag \\
    &\hspace{5mm}\times\left(\frac{5(1+3z^2)}{(3+5z^2)^2}\phi_0(z)+\frac{2(3+8z^2+5z^4)}{z(3+5z^2)^2}\phi^\prime_0(z)\right) \notag \\    
    &J^{\rm other}(z)=\frac{d}{dz}\Big(\frac{36\kappa z(1+z^2)^2\phi_0(z)^2}{3+5z^2}\Big) \label{W}
\end{align}

By comparing with Eqs.~\eqref{temporal} and ~\eqref{spatial}, the each glueball modes can be expressed as
\begin{align}
    &t_2(t)=\frac{1}{6L^7}\int_{-\infty}^\infty dz\sum^\infty_{n=1}\frac{\alpha^{\rm T}_n\Psi^{\rm T}_n(r^\prime)}{(m^{\rm T}_n)^2+\vec{\Delta}^2}J(z) \\
    &t_0(t)=\frac{1}{3L^7}\int_{-\infty}^\infty dz\sum^\infty_{n=1}\frac{\alpha^{\rm T}_n\Psi^{\rm T}_n(r^\prime)}{(m^{\rm T}_n)^2+\vec{\Delta}^2} \notag \\
    &\hspace{12mm}\times\left[J(z)+\frac{3}{2}J^{\rm S}(z)+\frac{3}{2}J^{\rm other}(z)\right] \\
    &s_0(t)=-\frac{1}{2L^7}\int_{-\infty}^\infty dz\sum^\infty_{n=1}\frac{\alpha^{\rm S}_n\Psi^{\rm S}_n(r^\prime)}{(m^{\rm S}_n)^2+\vec{\Delta}^2}J^{\rm S}(z)
\end{align}
Using these expressions, the GFFs can be identified as
\begin{align}
    &A(t)=6t_2 \label{A_t2} \\
    &D(t)= -2(t_2+t_0+s_0) \label{D_t2_t0_s0}
\end{align}
From these results, we observe that the form factor $A(t)$ is entirely saturated by the tensor glueball (${\rm T}_4(2^{++})$) contribution, whereas $D(t)$ receives contributions from all three glueball modes. 

\section{Numerical results}

We present numerical results for the pion’s GFFs and the associated spatial distributions in the chiral limit ($m_\pi=0$). 

To evaluate the infinite sums, we perform an expansion in powers of $\vec{\Delta}^2$ as
\begin{align}
    &\frac{6}{L^7}\int^\infty_Rdr^\prime\sum^\infty_{n=1}\frac{\alpha^{\rm T/S}_n\Psi^{\rm T/S}_n(r^\prime)}{(m^{\rm T/S}_n)^2+\vec{\Delta}^2}=\sum^\infty_{k=0}F_k^{\rm T/S}(r^\prime)(-\vec{\Delta}^2)^k, \notag \\
    &F^{\rm T/S}_k(r^\prime)=\int_{R}^{\infty} dr\frac{r^3}{L^3}\sum_{n=0}^{\infty} \frac{\Psi_{n}^{\rm T/S}(r)\Psi_{n}^{\rm T/S}(r^{\prime})}{(m_{n})^{2k}}. \label{expand}
\end{align}
From the eigenvalue equations \eqref{EVeq}, $F^{\rm T/S}_k(r)$ obeys the following recursive relations:
\begin{align}
    &\frac{1}{r^3L^4}\partial_r\big((r^7-rR^6)\partial_rF^{\rm T}_k\big)=-F^{\rm T}_{k-1}, \\
    &\frac{1}{r^3L^4}\partial_r\Big((r^7-rR^6)\Big(\partial_r+\frac{144r^5R^6}{(5r^6-2R^6)(r^6+2R^6)}\Big)F^{\rm S}_k\Big) \notag \\
    &\hspace{50mm}=-F^{\rm S}_{k-1},
\end{align}
with the normalization $F^{\rm T/S}_0=1$ from Eq.~\eqref{CR}.
The functions $F^{\rm T/S}_{k\neq0}(r)$ can be determined iteratively using the boundary conditions $\partial_r F^{\rm T/S}_k(r)\big|_{r\rightarrow R}=1$ and $F^{\rm T/S}_k(r)\big|_{r\rightarrow \infty}=0$.

\begin{figure}
    \includegraphics[scale=0.25]{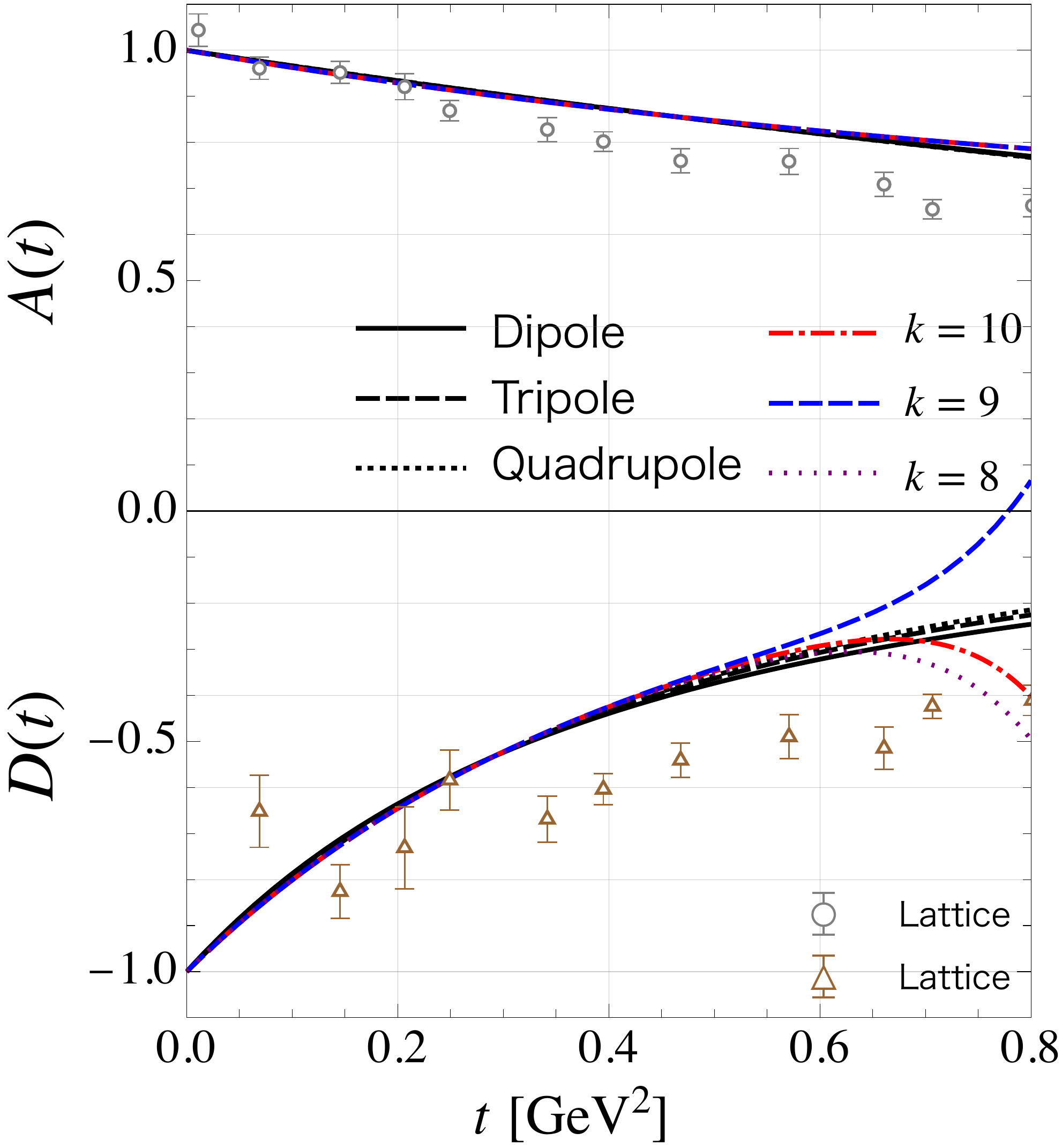}
    \caption{The pion GFFs $A(t)$ and $D(t)$, shown together with their multipole model fits and the lattice QCD results of Ref.~\cite{Hackett:2023nkr}.}
    \label{AtDt}
\end{figure}

We present the GFFs computed by truncating the series in Eq.~\eqref{expand} at $k=10$ in FIG.~\ref{AtDt}, along with the results truncated at $k=8$ and $k=9$, and, for comparison, the lattice QCD results of Ref.~\cite{Hackett:2023nkr}. 
The parameters are fixed as $\kappa =0.00745$ and $M_{\rm KK} =949  \ \rm{MeV}$, determined from the $\rho$ meson mass $776 \ {\rm MeV}$ and the pion decay constant $f_{\pi}=92.4 \ \rm{MeV}$~\cite{Sakai:2005yt}. While $A(t)$ shows good convergence up to $t \simeq 0.8 \ {\rm GeV^2}$ at $k=10$, $D(t)$ begins to diverge from around $t \sim 0.5 \ {\rm GeV^2}$. Since this model is expected to lose validity as $t$ approaches $M_{\rm KK}$, further extension is not required.


In the top-down holographic QCD, the magnitude of $D(t)$ decreases more rapidly than $A(t)$, in qualitative agreement with lattice QCD~\cite{Hackett:2023nkr}. We attribute this difference to the contribution of the ${\rm S}_4$ mode to $D(t)$, which is not present in the bottom-up framework of Ref.~\cite{Abidin:2008hn} because that model does not include an ${\rm S}_4$ fluctuation sourced by $\mathcal{T}_{\tau\tau}$.
Our analysis is performed in the large-$N_c$ and chiral limits, so a deviation from lattice QCD results is expected. In particular, in the large-$N_c$ regime, the exchange of flavor-singlet scalar mesons (e.g., the sigma meson) is suppressed by $1/N_c$, so the GFFs are dominated by scalar glueball exchange. As a result, the pion radius is reduced, i.e., $A(t)$ exhibits a slower fall-off. In contrast, $D(t)$ decreases faster than in lattice QCD result, which we attribute to the relatively small mass of the ${\rm S}_4(0^{++})$ state that contributes only to $D(t)$. As noted in footnote~\ref{footnote2}, this state has a lower mass and a larger decay width than the lightest scalar glueball in QCD, and may therefore be a holographic artifact.

In addition, Fig.~\ref{AtDt} shows fits to the multipole ansatz
$F(t)=F(0)/(1+t/\Lambda^{2})^{m}$ with $F(0)=A(0)$ or $D(0)$ for
$m=2,3,4$ (dipole, tripole, quadrupole), plotted as curves together with the GFFs results.
The corresponding best-fit values of $\Lambda^{2}$ for $A(t)$ and $D(t)$ at each $m$ are listed in Table~\ref{tab1}.
\begin{table}[t!]
\centering
\caption{Multipole-fit parameters $\Lambda^{2}$ for $A(t)$ and $D(t)$ at $m=2,3,4$.}
\begin{tabular}{lcccccc}
\hline\hline
$m$ & $\Lambda^2/{\rm GeV^2}$ for $A(t)$ & $\Lambda^2/{\rm GeV^2}$ for $D(t)$ \\
\hline
$2$ (Dipole) & $6.36$ & $0.87$ \\
$3$ (Tripole) & $9.67$ & $1.38$ \\
$4$ (Quadrupole) & $12.98$ & $1.89$ \\
\hline\hline
\end{tabular}
  \label{tab1}
\end{table}
This multipole ansatz satisfies the following Fourier transform with the modified Bessel function of the second kind $K_m$:
\begin{align}
    \int\frac{d^2\vec{\Delta}_\perp}{(2\pi)^2}\frac{e^{-i\vec{\Delta}_\perp\cdot\vec{x}_\perp}}{(1+\vec{\Delta}^2_\perp/\Lambda^2)^m}=\frac{\Lambda}{\pi x_\perp}\left(\frac{\Lambda x_\perp}{2}\right)^m\frac{K_{m-1}(\Lambda x_\perp)}{(m-1)!}, \notag
\end{align}
which provides qualitative insight into how spatial densities behave, following the approach of, e.g., Refs.~\cite{Lorce:2018egm,Freese:2021czn}.

\begin{figure}
    \includegraphics[scale=0.38]{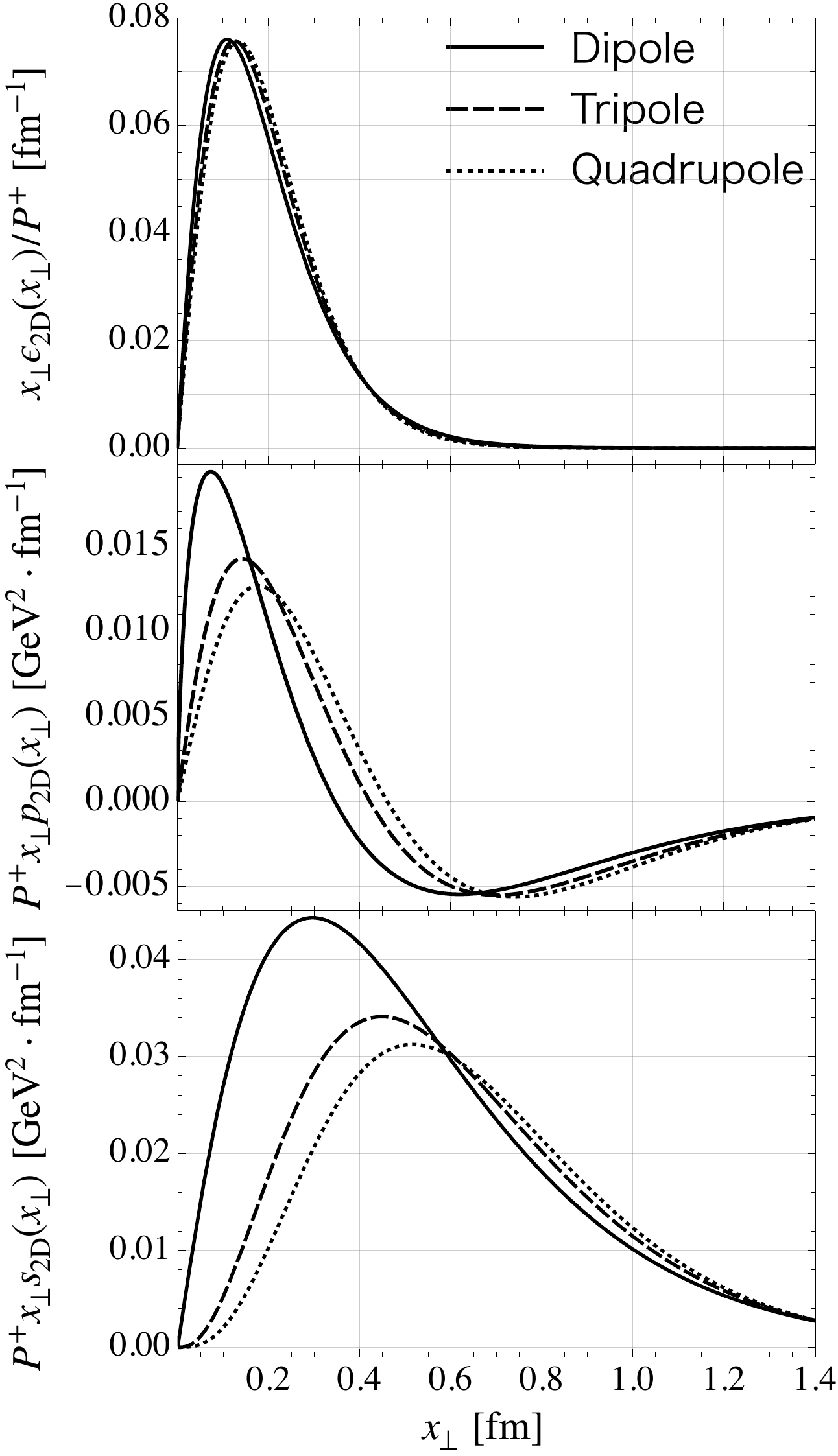}
    \caption{Energy density $\epsilon_{\rm 2D}(x_\perp)$, pressure $p_{\rm 2D}(x_\perp)$ and shear force distributions $s_{\rm 2D}(x_\perp)$.}
    \label{FIG_2_density}
\end{figure}

\begin{figure}
    \includegraphics[scale=0.17]{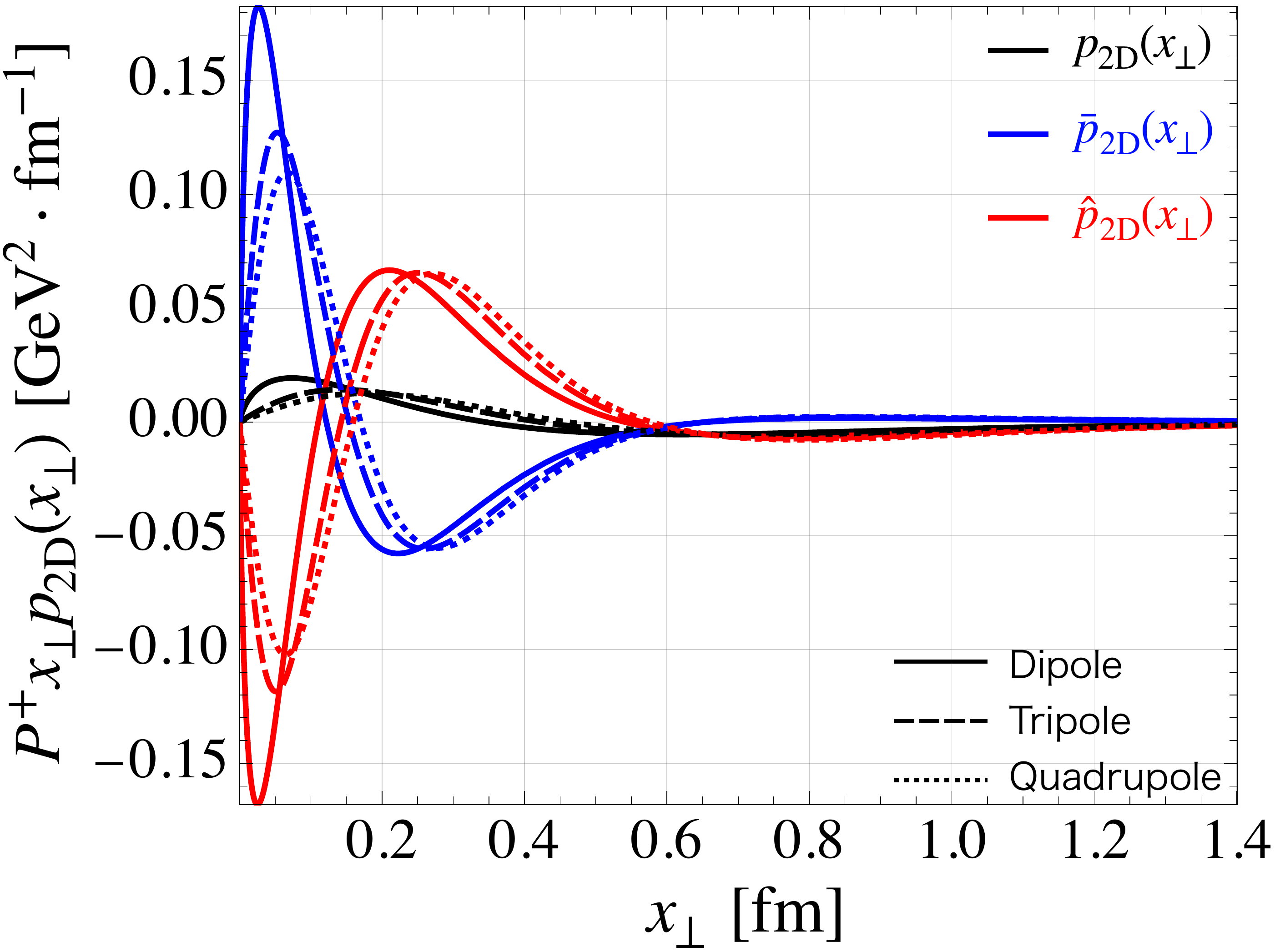}
    \caption{Pressure decomposition.}
    \label{pressure_decomp}
\end{figure}

Using this multipole model, we evaluated the transverse energy density $\epsilon_{\rm 2D}(x_\perp)$, pressure $p_{\rm 2D}(x_\perp)$, and shear force $s_{\rm 2D}(x_\perp)$ based on Eqs.~\eqref{2Depsilon}–\eqref{Spherical2Dstress}, and present the results in Fig.~\ref{FIG_2_density}. As seen from these equations, $\epsilon_{\rm 2D}$ is uniquely associated with the tensor glueball, in contrast to $p_{\rm 2D}$ and $s_{\rm 2D}$ which receive contributions from all three glueball modes. 

Next, we present in Fig.~\ref{pressure_decomp} the decomposition of the pressure based on Eq.~\eqref{decomp_pressure}. Near the center, $\bar{p}_{\rm 2D}$ generates an outward pressure, whereas $\hat{p}_{\rm 2D}$ yields an inward, confining pressure. Although the magnitudes of each contribution are individually large, they mostly cancel each other, resulting in a total pressure whose magnitude is suppressed to roughly $1/10$. In the intermediate region, a sign flip is observed for both $\bar{p}_{\rm 2D}$ and $\hat{p}_{\rm 2D}$, which is a novel behavior not captured in previous analyses for nucleons in the instant form~\cite{Fujii:2025aip}. 
In contrast, near the pion surface, the pressure components flip their signs again, and as discussed in Ref.~\cite{Fujii:2025aip}, $\hat{p}_{\rm 2D}$ induced by the trace anomaly provides a confining pressure. Furthermore, from Eqs.~\eqref{decomp_pressure}, \eqref{A_t2} and \eqref{D_t2_t0_s0}, it is found that the contribution from the ${\rm T}_4(2^{++})$ mode to the trace anomaly cancels out exactly, indicating that the scalar glueball modes ${\rm T}_4(0^{++})$ and ${\rm S}_4(0^{++})$ are responsible for mediating the effects of the trace anomaly.

The fact that the pressure originating from the scale anomaly provides the confining pressure has been confirmed in a model-independent manner in several previous studies~\cite{Liu:2023cse,Ji:2025gsq,Fujii:2025pkv,Ji:2025qax}. Furthermore, in the large-$N_c$ limit, the exchange of flavor-singlet scalar mesons is suppressed by $1/N_c$, so the GFFs are dominated by scalar glueball exchange. This has been discussed in bottom-up holographic QCD for the nucleon~\cite{Mamo:2021krl} and in the instanton liquid model for both the nucleon~\cite{Liu:2024rdm} and the pion~\cite{Liu:2024jno}. In particular, Refs.~\cite{Liu:2024rdm,Liu:2024jno} demonstrate the scalar glueball dominance in the matrix element of the scale anomaly $\langle \hat{T}^{\mu\nu} \rangle$, and confirm that these results are consistent with recent lattice QCD calculations~\cite{Wang:2024lrm}. These studies in the large-$N_c$ limit of QCD support the universality of the phenomenon that we have found in the top-down holographic QCD approach in this paper.

Finally, based on the tripole model fit, we calculate the mass radius $\braket{x_\perp^2}_{\rm mass}$ and the mechanical radius $\braket{x_\perp^2}_{\rm mech}$ as
\begin{align}
    &\sqrt{\braket{x_\perp^2}_{\rm mass}}=0.23 \ {\rm [fm]}, \notag \\ &\sqrt{\braket{x_\perp^2}_{\rm mech}}=0.50 \ {\rm [fm]}, \notag 
\end{align}
respectively. As discussed above, within the top-down holographic QCD framework, the former tends to be underestimated while the latter tends to be overestimated.

\section{Summary and outlooks}

In this work, we have investigated the two-dimensional transverse density distributions in the LF form, derived from the GFFs of the pion, using a top-down holographic QCD approach. Following Ref.~\cite{Fujii:2025aip}, we performed a trace–traceless decomposition of the EMT to isolate the contribution of the trace anomaly to the pressure distribution. As a result, we have identified the scale-anomaly-driven confining pressure—previously confirmed in the nucleon within the instant-form—as also present in the pion on the LF form. The emergence of this same mechanism across distinct dynamical forms (instant vs.\ LF) and hadron types (nucleon vs.\ pion) suggests that the dominance of the scale anomaly in generating confining pressure is a universal feature of hadron dynamics. Moreover, we found that among the contributing glueball states ${\rm T}_4 (2^{++})$, ${\rm T}_4 (0^{++})$, and ${\rm S}_4 (0^{++})$, the scalar glueballs are responsible for the generation of the confining pressure.

On the other hand, the positive contribution to the pressure induced by the trace anomaly in the intermediate region appears to be a notable feature of the Nambu–Goldstone boson nature of the pion. This is supported by the fact that each component of the pressure decomposition satisfies the von Laue condition individually, i.e., $\int d^2x_\perp \bar{p}_{\rm 2D}(x_\perp) =m_\pi^2/4= 0$ and $\int d^2x_\perp \hat{p}_{\rm 2D}(x_\perp) =- m_\pi^2/4=0$. 
It would be of great interest to further explore this characteristic behavior by incorporating current quark mass effects following Ref.~\cite{Casero:2007ae,Hashimoto:2007fa,Bergman:2007pm,Dhar:2007bz,Hashimoto:2008sr,Dhar:2008um,McNees:2008km}, which we leave for future work. This approach, when applied to different hadrons such as the nucleon, can further clarify which features of the stress distribution are universal and which arise from the intrinsic properties of each hadron.

\section*{Acknowledgments}

We are grateful to the authors of Ref.~\cite{Hackett:2023nkr} for generously supplying the data tables utilized in FIG.~\ref{AtDt}. 
This work was supported in part by the Japan Society for the Promotion of Science (JSPS) KAKENHI (Grants No. JP24K17054, JP24KJ16200) and the COREnet project of RCNP, Osaka University.


\bibliography{ref}

\end{document}